\documentstyle[prl,twocolumn,aps]{revtex}
\title{\bf Quasiparticles and the quantum fluctuations of local observables}
\author{S. Ying}
\address{CCAST (World Laboratory), P. O. Box 8730, Beijing 100080, 
         People's Republic of China\\ and \\ Physics Department, Fudan
        University, Shanghai 200433, People's Republic of China\thanks{The 
         present address.}}
\date{\today}
\tighten
\narrowtext
\draft
\begin{document}
\maketitle

\begin{abstract}
   The role that quasiparticles play in a strong interaction system with
   spontaneous symmetry breaking is examined. We find, using a non-
   perturbative cluster decomposition method, that the quasiparticles do
   not saturate the physical local observables at small distances. The
   fermion number density serves as a clearcut example. A component due to 
   localized random quantum fluctuations of the order parameter(s) in the
   vacuum state and the contributions of ``quasiparticles'' corresponding to 
   other local minima of the effective potential is needed. At large 
   distances, the ordinary quasiparticle picture emerges in the response of 
   the system to classical background fields but the above mentioned component
   acts as a source for them.
\end{abstract}

  In the self-consistent mean field and Hartree--Fock approximations, the
behavior of a quasiparticle is similar to an ordinary particle. A particle in
a free or weakly interacting theory has definite mass; observables of the
theory are saturated by the contributions of these particles. As a
consequence, the fermion number density of a system with massive fermions is
non-zero only after the chemical potential $\mu$ exceeds the mass of the
lightest fermions in the system. It is expected that the quasiparticles do
the same at the mean field level. 

Quasiparticles are however not particles. The essential difference between
the mass of a particle and that of a quasiparticle lies in the fact that the
mass of a particle is a macroscopic parameter of the theory, with
no quantum fluctuations, and the ``mass'' of the quasiparticle, being
proportional to the order parameter, is a local variable. Such a local
variable contains quantum fluctuations that do not decouple in the
thermodynamical limit. As a result the fermion number density is not
saturated solely by the contributions of the quasiparticles but contains a
component due to local quantum fluctuations. This component, being
non-propagating and discorrelated with each other in different spacetime
region, is called the dark component.

    Recent lattice studies of the spontaneous chiral symmetry breaking in QCD
at finite density found the emergence of baryon number at chemical potential
below the expected threshold for particle production \cite{steph,shuyak} even
in the non-quenched approach \cite{shuyak}. The origin of this effect is not 
yet well understood. It is found here that the dark component is responsible 
for such a behavior.

For simplicity, the two flavor half bosonized NJL model is used for our
investigation. The model Lagrangian density with a chemical potential $\mu$
is
\begin{eqnarray}
{\cal L} &=& \bar\psi\left (i\rlap\slash\partial + \mu\gamma^0 -\sigma -
i\gamma^5 \vec{\tau}\cdot\vec{\pi} \right )\psi - {1\over 2 G_0} \left
(\sigma^2 + \vec{\pi}^2 \right ), \label{Model-L}
\end{eqnarray}
where $G_0$ is the coupling constant, $\sigma$ and $\vec{\pi}$ are auxiliary
fields and $\vec{\tau}$ is a set of three Pauli matrices. It has a
$SU(2)_L\times SU(2)_R$ chiral symmetry.

After performing the integration over the fermion fields $\psi$ and
$\bar\psi$, the generating functional $W[J]$ can be written as
\begin{eqnarray}
    e^{W[J,\mu]} &=& \int D[\sigma,\vec{\pi}]
     e^{iS_{eff}[\sigma,\vec{\pi},\mu] + i \int d^4x f\cdot J},
\label{Gen-Func2}
\end{eqnarray}
where ``$J$'' and ``$f$'' represent, collectively, the external fields and
auxiliary fields respectively. The effective action $ S_{eff}$ is given by
\begin{eqnarray}
S_{eff}[\sigma,\vec{\pi},\mu] &=& - i{1\over 2} Sp Ln
    S_F^{-1}[\sigma,\vec{\pi},\mu] S_F[0,0,0]+{1\over 2G} \int d^4 x \left (
            \sigma^2 + \vec{\pi}^2 \right),
\label{Seff}
\end{eqnarray}
where $Sp$ denotes the functional trace. The operator
$iS_F^{-1}[\sigma,\vec{\pi},\mu]= i\rlap\slash\partial+\mu\gamma^0 -\sigma -
i\gamma^5 \vec{\tau}\cdot\vec{\pi}$ is the inversed propagator of the
fermions in the background auxiliary fields $\sigma(x)$ and $\vec{\pi}(x)$.

Let us define $\rho[\sigma,\vec{\pi},\mu;x]\equiv \mbox{Tr}\gamma^0
\mathopen{\langle x\,|}S_F[\sigma,\vec{\pi},\mu] \mathclose{|\,x\rangle} $,
where the trace ``Tr'' is over the internal degrees of freedom of the
fermions. The fermion number density of the model in a state $
\mathclose{|\,\psi\rangle} $ determined by the initial and final
configurations of the auxiliary fields is
\begin{eqnarray}
\mathopen{\langle \psi\,|}\widehat\rho(x) \mathclose{|\,\psi\rangle}
&=& {1\over Z}\int D[\sigma,\vec{\pi}]\rho[\sigma,\vec{\pi},\mu;x]
e^{iS_{eff}[\sigma,\vec{\pi},\mu]},
\label{rho-general}
\end{eqnarray}
where $Z= \int D[\sigma,\vec{\pi}]e^{iS_{eff}[\sigma,\vec{\pi},\mu]}$. The
Minkowski spacetime is not suitable to study the properties of the vacuum
state using Eq. \ref{rho-general} since the initial and final auxiliary field
configurations are not specified. For a given set of configurations, the
contributing intermediate states are not merely the vacuum states but are
contaminated by the excited states. The usual procedure to project out the
contributions of the vacuum state is go to the Euclidean spacetime in which
the vacuum state has lowest energy.

In the mean field approximation, the vacuum phase is determined by minimizing
the effective potential $ V_{eff}(\sigma,\mu) = -
S_{eff}[\sigma,0,\mu]/\Omega$ with $\Omega\to\infty $ the spacetime volume of
the system, $\sigma$ spacetime independent and $\vec{\pi}$ assumed zero. Due
to the chiral symmetry, the assumption $\vec{\pi}=0$ does not loss
generality. The phase of the system is determined by the condition $\delta
V_{eff}(\sigma,\mu )/\delta\sigma =0$. The solution for $\bar\sigma$ is
non-zero after the coupling constant $G_0$ is greater than a critical value
$G_{0c}$. A non-vanishing $\bar\sigma$ generates an effective mass for the
fermions to form quasiparticles.

The mean field fermion number density  for the vacuum is
obtained from Eq. \ref{rho-general} by ignoring the functional integration
over $\sigma$ and $\vec{\pi}$ and let $\sigma=\bar\sigma$. The result is
\cite{FD-pap}
\begin{eqnarray}
\bar\rho_{MF} &=&
\rho(\bar\sigma,0,\mu;x) = \theta(\mu-\bar\sigma)\left
[\mu^2-\bar\sigma^2\right ]^{3/2},
\label{rhoMF}
\end{eqnarray}
 which is non-zero only when $\mu>\bar\sigma$ just like the fermion number
 density of free massive particles with mass $m=\bar\sigma$. Therefore, the
 quasiparticle contributions saturate the fermion number density in the mean
 field approximation.

The contributions of quantum fluctuations around the mean field $\bar\sigma $
are formally included in Eq. \ref{rho-general}. The results are commonly
expressed as loop corrections to the fermion number density vertex. It is not
attempted here.

Instead of performing a loop expansion computation of the fermion number
density, we evaluate non-perturbatively the effects of the quantum
fluctuations by ``doing'' the path integral.

To proceed, the system under consideration is first putted in a Euclidean
spacetime box of length $L$ with periodic boundary conditions at its
surfaces. The thermodynamical limit is defined as the limit of $L\to\infty$.
In the thermodynamical limit, the extremal configuration dominates the path
integral among those configurations of $\sigma$ and $\vec{\pi}$ that give
divergent action in the thermodynamical limit. The contributing finite action
quantum fluctuation configurations, the number of which is proportional to
the spacetime volume $\Omega=L^4$, are further classified into two
categories: 1) correlated localized configurations, which are defined as the
ones that approaches to the mean field configurations in the spacetime
infinity and 2) correlated extended configurations, which are the ones that
remain different from the mean field configurations at the spacetime
infinity.

For a given system, whether or not a configuration is an correlated extended
configurations or is of localized ones\footnote{In the sense that it can be
decomposed into localized ones.} is determined by dynamics. The on-shell
amplitudes, are solutions of the ``classical equation of motion'' in the
Euclidean spacetime
\begin{eqnarray}
{\delta S^E_{eff}[f]\over\delta f(x) } &=& 0,
\label{Eq-onshell}
\end{eqnarray}
with $S^E_{eff}$ the Euclidean effective action, $f$ representing $\sigma$ or
$\vec{\pi}$ fields. The superscript $E$ shall be suppressed in the following.
The set of the extended solutions to Eq. \ref{Eq-onshell} are correlated
ones. Albeit there are plenty of extended on-shell amplitudes in the
Minkowski spacetime, there is no known one in the Euclidean one. We shall
assume the absence of them. The degree of correlation of an arbitrary
extended configuration at different spacetime points is determined by the
degree of their deviation from the extended on-shell amplitudes for
propagating excitations of the system.

The correlation between the off-shell configurations at two different space
time points decreases exponentially. For example, consider the field--field
correlations or propagators
\begin{eqnarray}
    \mathopen{\langle 0\,|}T \widehat f(x_1) \widehat f(x_2) 
    \mathclose{|\,0\rangle} &=&
    \int {d^4 p\over (2\pi)^4} e^{-ip\cdot (x_1-x_2)} {1\over p^2+m^2}
\label{E-fprop}
\end{eqnarray}
with $T$ denoting time ordering. The off-shellness of these configurations is
measured by their mass $m$. The correlation of these configurations at two
different location ${x}_1$ and $ {x}_2$ (in the Euclidean spacetime)
decreases as fast as $\exp(-| {x}_1- {x}_2| m)$. So they can be decomposed
into superposition of localized ones with a size of order $1/m$. For these
set of configurations, one can divide the spacetime into cells with a
dimension sufficiently larger than their correlation length. Then the
contribution of this set of field configurations to the partition functional
Eq. \ref{Gen-Func2} can be cluster decomposed to
\begin{eqnarray}  Z[J] &\approx & \prod_k
 z_k[J],\hspace{0.4cm} W[J] \approx \sum_k  w_k[J]
     \label{Cluster-Decomp}
\end{eqnarray}
with $ w_k[J]\equiv \ln  z_k[J]$, $ z_k[J]$ and $ w_k[J]$ the corresponding
partition functional of the kth cell. The full partition of the kth cell is
\begin{eqnarray}  z_k[0] &\sim & \int D_k[f] e^{-S^{(k)}_{eff}(\bar
      f+f)},
\label{Cell-Pint}
\end{eqnarray}
where $S^{(k)}_{eff}[f]$ is the effective action of the kth cell and the path
integration of $f$ is over those ones that equals to the mean field value
$\bar f$ outside of the kth cell but with arbitrary amplitudes inside the
finite volume. For a given theory, instead of arbitrary division of the
spacetime, it is expected that there is an optimal one with minimum volume
$\varpi$ for each cell and yet has an error below a predetermined one. We
shall assume that such an optimal division of the spacetime into cells has
already been found in the following discussion.

Let us evaluate the contributions of the uncorrelated localized quantum
fluctuations of the $\sigma$ and $\vec{\pi}$ fields to the vacuum fermion
number density using Eqs. \ref{rho-general}, \ref{Cluster-Decomp} and
\ref{Cell-Pint} in the Euclidean spacetime.

In the phase where the chiral $SU(2)_L\times SU(2)_R$ symmetry is
spontaneously broken down, the ``chiral angle'' variable represented by
$\vec{\pi}$ in the phase where $ \mathopen{\langle 0\,|} \vec{\pi}
\mathclose{|\,0 \rangle}=0$ becomes massless following the Goldstone theorem.
The correlation length of the $\vec{\pi}$ field becomes divergent in the
chiral symmetrical limit. Therefore the field configurations of the Goldstone
boson degrees of freedom contain the dominating on-shell components that can
be included by doing a loop expansion as usual. Such a loop expansion
contains no infrared divergences. Because the fermion number density $\rho
(\sigma,\vec{\pi},\mu;x)$ under study is chiral symmetric, which means that
it does not depend on a spacetime independent global ``chiral angle''; it
depends only on the derivatives of the ``chiral angle'' variables. The
absence of a dependence  of the fermion number density on a global ``chiral
angle'' guarantees the absence of the infrared divergences in the quantum
corrections due to the Goldstone bosons. The configurations of the ``chiral
angle'', being on the edge of their shell and extended configurations in
nature, are uniform and infinitesimal in amplitudes since it contains no
infrared divergences and has a number of distinct modes proportional to the
volume $\Omega $ of the system. They can not modify the qualitative features
of the quasiparticles. So, the $\vec{\pi}$ variable in the vacuum fermion
number density can be eliminated. It is treated as zero in the following
discussions.

The quantum fluctuation in the ``mass'' term, namely the chiral radius or
order parameter represented by $\sigma$ (when $ \mathopen{\langle
0\,|} \vec{\pi} \mathclose{|\,0\rangle}=0$) has different 
characteristics due to the fact that it contains no on-shell Euclidean 
configurations. These
off-shell configurations have only short range correlations in spacetime.

   If the spacetime is divided into cells with their dimension optimally
determined, then the path integration within each cell can be done
independently. This gives us a cluster decomposed partition functional of the
form given by Eq. \ref{Cluster-Decomp} with the partition functional for each
cell computed independently of each other.

The cluster decomposition property of the partition functional of the system
reduces the full fermion number density of the vacuum given by Euclidean form
of Eq. \ref{rho-general} to
\begin{eqnarray}
       \rho_{vac}  &=& {1\over
       z_k[0]}\int D_k[\sigma'] \rho[\bar\sigma+\sigma',0,\mu;x]
       e^{-S^{(k)}_{eff}[\bar\sigma+\sigma',\mu]} \label{rho-decomp}
\end{eqnarray}
where the cell labeled by k is the one that contains the spacetime point $x$,
$\int D_k[\sigma']$ denotes integration over field configurations that
approach the mean field value outside the cell and $S^{(k)}_{eff}$ is the
Euclidean effective action of the cell.

Eq. \ref{rho-decomp} is still too complicated to evaluate analytically. We
make a further simplification by assuming that the functional integration of
$\sigma'$ within a spacetime cell can be replaced by an ordinary integration
over the spacetime averaged value of $\sigma'$ within that cell. It can be
achieved by keeping the average of $\sigma'$ fixed while integrate over the
rest degrees of freedom. The non-trivial part of it is in the assumption that
after eliminating the rest of the degrees of freedom, the resulting
effective potential $V_{eff}$ remains, at least in form, the same as the
original one. This procedure is in the same spirit as the renormalization
group analysis. In this way, Eq. \ref{rho-decomp} reduces to
\begin{eqnarray}
       \rho_{vac} &=& {1\over
       z[0]}\int^\infty_{-\infty}
       d\delta\sigma \rho[\bar\sigma+\delta\sigma,0,\mu;x]
       e^{-\varpi V_{eff}(\bar\sigma+\delta\sigma,\mu)},
\label{rho-decomp1}
\end{eqnarray}
where $\delta\sigma=<\sigma'>$ is the spacetime average of $\sigma'$ within
the cell and the same reduction is also made to $z[0]$. The finiteness of
$\varpi $ result in different qualitative behavior for $ \rho_{vac}$ as a
function of $\mu$. To explicitly see the difference, let us expand $V_{eff}$
around $\bar\sigma$, keeping only the leading quadratic term
\begin{eqnarray}
   V_{eff}(\bar\sigma+\delta\sigma) &=& V_{eff}(\bar\sigma) +
                              A(\bar\sigma) \delta\sigma^2 + \ldots
\end{eqnarray}
and using Eq. \ref{rhoMF} for the trace term. If $\varpi A$ is sufficiently
large, the result can be written as
\begin{eqnarray}
\rho_{vac} &=& {2\over \pi^2} \sqrt{\varpi
 A\over\pi}\int_{-\mu-\bar\sigma}^{\mu-\bar\sigma} d\delta\sigma
        e^{-\varpi A \delta\sigma^2} \left
        [\mu^2-(\bar\sigma+\delta\sigma)^2\right ]^{3/2}.
\end{eqnarray}
It is clearly non-zero for any finite $\mu$ below $\bar\sigma$ since $A$ is
finite except in the limit of $G_0\to 0$, which represents the free field
case. The dependence of $\rho_{vac}^{1/3}$ on $\mu$ for a set of different
values of $\alpha=\varpi A$ is plotted in Fig. 1. Instead of a sharp rise in
$\rho^{1/3}_{vac}$ at $\mu = \bar\sigma$, $\rho$ is non-vanishing all the way
to $\mu=0$. This component of the fermion number density can certainly not be
attributed to the contributions of the quasiparticles. We find therefore that
there is a dark component for the fermion number density that can not be
accounted for by the quasiparticle contributions.

The NJL model has only one non-trivial vacuum phase since there is one
independent absolute minimum in its effective potential. If the effective
potential of the system contains a second local minimum with a higher energy
density than the absolute minimum, then there are contributions from the local
minimum to the fermion number density. The existence of such a contribution
is another direct consequences of the cluster decomposability of the
partition functional of the system. Models with a second minimum are studied
in the literature. Those that have only one order parameter are represented
by the Friedberg--Lee model \cite{FL-model} in which the two minima of the
effective potential correspond to confinement and deconfinement phase of the
model. Those that has two order parameters are introduced in Refs.
\cite{pPaps,pPaps1} in which the $\beta$ and $\omega$ phases of the massless
fermion system are studied.

In the presence of a higher virtual phase that is separated from the actual
phase of the system by a potential barrier, the fermion number density of the
system is saturated by both the quasiparticles of the first phase and those
ones of the second virtual phase together with their corresponding dark
component discussed above. It has a form
\begin{eqnarray}
\rho_{vac} &=& {\rho_{vac_1} + e^{-\varpi
  \Delta} \rho_{vac_2}\over 1 + e^{-\varpi \Delta}},
\label{2phase-rho}
\end{eqnarray}
where $\Delta=V_{eff}^{(2)}-V_{eff}^{(1)}>0$ is the difference in energy
density between the virtual phase and the actual phase, $\rho_{vac_1}$ and
$\rho_{vac_2}$ are the vacuum fermion number density of the actual vacuum
phase and that of the virtual vacuum phase respectively and $\varpi$ is the
optimal volume of the spacetime cell between which the order parameters of
the system are discorrelated. The contributions of the virtual phase to the
fermion number density or to any local physical observables are
non-perturbative effects. Attempts had been made in Refs.
\cite{pPaps1,JPaps} to search for possible other virtual phases 
\cite{pPaps,pPaps1} of the strong interaction vacuum.

   The fermion number density discussed above are defined on a spacetime
point. Such a precision is non-achievable in realistic observations. The
physical observables can be represented by a coarse-grained averaging of the
form $ \bar\rho_{vac}={\Delta N_{vac} /\Delta\Omega}$ with $\Delta\Omega$ the
smallest volume in spacetime that the observation apparatus can resolve and
$\Delta N_{vac}$ the average number of fermions {\em coherently}
produced by an external
classical field $K$ within that volume. $\Delta N_{vac}$ is not in general
identical to $\int_{\Delta\Omega} d^4 x \rho_{vac}$ when 
$\Delta\Omega>>\varpi$. It is obtained from
the partition functional $W[J,\mu]$ by adding the external field $K$ to $\mu$.
$K$ has a constant non-zero value only within spacetime volume $\Delta
\Omega$, then $\Delta N_{vac} = \partial W[0,\mu+K]/\partial iK |_{K=0}$. In
case when $\Delta \Omega << \varpi$, namely the precision of the observation
is much higher than the correlation length of the order parameter, the
observed fermion number density $\bar\rho_{vac}$ behaves in the same way as
$\rho_{vac}$. On the other hand, if $\Delta\Omega >> \varpi$, then the
smallest spacetime cell that contributes to $\Delta N_{vac}$ is a region of
volume $\Delta\Omega$ rather than $\varpi$. In this case, $\bar\rho_{vac}$ is
obtained from $\rho_{vac}$ by substituting $\Delta\Omega$ for $\varpi$. The
effects of the dark component in the observed fermion number density are
reduced as a result. When the resolution is sufficiently low, which means
$\Delta \Omega A >> 1$, the quasiparticle picture with only one vacuum ,
namely the actual one, reemerges in the response of the system to $K$.
Therefore the dark component of the fermion number density is of transient
nature.

On the other hand, the $K$ field radiated by the vacuum fermion number density
is of the form $K(x)=\int d^4 x'G(x,x')\rho_{vac}(x')$ with $G(x,x')$ the 
Green function for $K(x)$. It is $\rho_{vac}$ that is the source for $K(x)$ 
rather than $\bar\rho_{vac}$ no matter how slow the resulting $K(x)$ varies
in spacetime. Thus the effects of the dark component are indirectly observable
even in low energy processes.

Albeit the fermion number density is discussed, the results of the letter are
also applicable to other local observables like the energy density. The
finding may have implications on the dark matter problem in cosmology since
it implies that the apparent matterless space at the macroscopic level may
reveal itself of matter effects in low energy gravitational processes even
when $\mu$ is below the baryonic particle production threshold and the $\mu=
0$ energy density subtracted. This is because gravitational fields couple 
locally to the source matter fields that contain the localized dark component.

In summary, it is found that the quasiparticle picture is insufficient in
describing quantum processes of strong interaction at both small and large 
distance scales. The random quantum fluctuations of order parameters of the 
system can change the behavior of the system qualitatively. The finding may 
be relevant to our understanding of the dark matter problem in astronomy.
Other implications in hadron physics are also worth exploring in the future.

\section*{Figures}

\noindent
{\bf Fig. 1} The dependence of $\rho_{vac}^{1/3}$ on the chemical potential
$\mu$. Here $\alpha =\varpi A$. The unit for the dimensional quantities are
$GeV$. Solid lines represent the case of free theory with mass 0 and 0.5
respectively. Other lines represent the results for the massless NJL model
with different strength of local fluctuations characterized by the $\alpha$
of the order parameter.

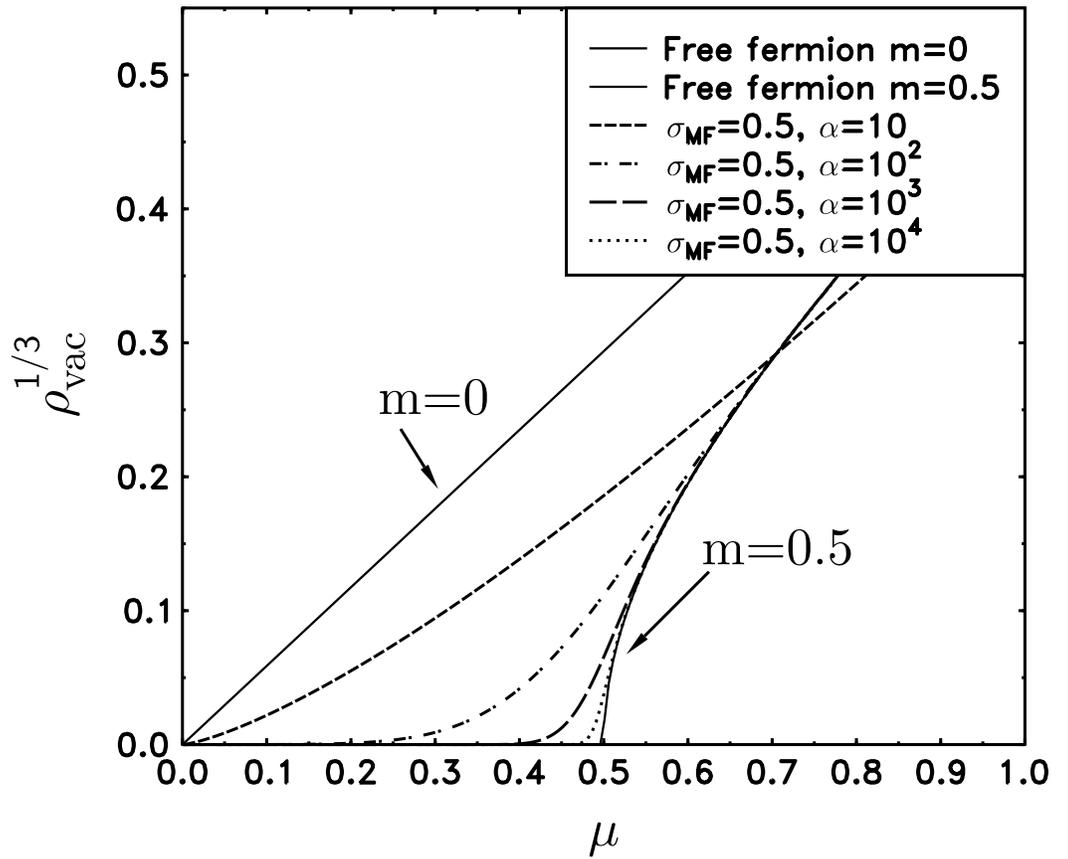
\end{document}